\documentclass[aps,prb,twocolumn,superscriptaddress]{revtex4-2}
\usepackage[utf8]{inputenc}
\usepackage{amsmath}
\usepackage{graphicx}
\usepackage{subfigure}
\usepackage{xcolor}
\usepackage{amsfonts}
\usepackage{booktabs}
\usepackage{bigstrut,multirow,rotating}
\usepackage{appendix}
\usepackage{svg}
\usepackage{hyperref}
\hypersetup{
  colorlinks=true,
  citecolor=magenta,
  linkcolor=red,
  urlcolor=cyan}

\begin{document}

\title{Phase diagram of a square lattice model of XY Spins with direction-dependent interactions}
\author{Fan Zhang}
\affiliation{Department of Physics, Beijing Normal University, Beijing 100875, China}

\author{Wenan Guo}
\email{waguo@bnu.edu.cn}
\affiliation{Department of Physics, Beijing Normal University, Beijing 100875, China}
\affiliation{Key Laboratory of Multiscale Spin Physics, Ministry of Education, Beijing 100875, China}
\author{Ribhu Kaul}
\email{ribhu.kaul@psu.edu}
\affiliation{Department of Physics, The Pennsylvania State University, University Park PA-16802, USA}
\date{October 2022}

\begin{abstract}
   
    We study a generalization of the well-known classical two-dimensional square lattice compass model of XY spins (sometimes referred to as the 90$^\circ$ compass model), which interpolates between the XY model and the compass model. Our model possesses the combined $C_4$ lattice and spin rotation symmetry of the compass model but is free of its fine-tuned subsystem symmetries.  Using both field theoretic arguments and Monte Carlo simulations, we find that our model possesses a line of critical points with continuously varying exponents of the Ashkin-Teller type terminating at the four-state Potts point.  Further, our Monte Carlo study uncovers that beyond the four-state Potts point, the line of phase transition is connected to the lattice-nematic Ising phase transition in the square lattice compass model through a region of first-order transitions.
\end{abstract}

\maketitle

\section{Introduction}

Spin models which have bond-direction dependent interactions, also called ``compass models," have provided a novel perspective to important problems such as classical frustration 
 and the emergence of quantum spin liquids\cite{compass_rev}
The most famous of these, the Kitaev honeycomb model,  displays both rich classical frustration~\cite{baskaran2008} and anyonic excitations in the quantum limit\cite{Kitaev2005}. Recent interest in the study of spin-orbit coupled Mott materials
\cite{Jackeli-Mott} has provided a fresh impetus to understand such models in various contexts that arise in experiments~\cite{Hermanns-Kitaev}.

Perhaps one of the simplest examples of such a bond-direction dependent interaction is in the classical square lattice compass model (CM)\cite{compass_rev, Mishra},
\begin{equation}
H_{\rm CM} = - J \sum_i (S^x_i S^x_{i+\hat x} +S^y_i S^y_{i+\hat y})
\label{H_compass}
\end{equation}
defined with a two-component classical field $\vec S_i = (S_i^x,S_i^y)$ on the sites $i$ of the square lattice. This model has some unusual symmetries as compared to the standard internal and lattice symmetries of statistical mechanics models. First, note that a site-centered rotation of the lattice by $\pi/2$ is not a symmetry. Neither is an internal rotation of the $\vec S$ by $\pi/2$, {\em i.e.} $(S_i^x, S_i^y)\rightarrow (-S_i^y, S_i^x)$ but combining these two lattice and internal operations together results in a symmetry for the system. We shall call this combined space-internal rotation, ${\cal C}_4$,  which will play an important role in our work. Such operations arise naturally in spin-orbit coupled Mott insulators in which the spin and space must be rotated together in the implementation of physical space group transformations. In addition, the model Eq.~(\ref{H_compass}) has another family of striking ``sub-system" symmetry operations in which $S_i^x\rightarrow -S_i^x$  on any one row and $S_i^y\rightarrow -S_i^y$ on any one column. These operations clearly forbid traditional spin-spin long-range order~\cite{batista05_gls}. Numerical work has established that the CM has a transition into a lattice nematic that breaks the rotation symmetry, an Ising critical point separates the ordered and disordered phases.\cite{Mishra, Wenzel_2010}
\begin{figure}[t]
\centering
\setlength{\abovecaptionskip}{20pt}
\includegraphics[angle=0,width=0.48\textwidth]{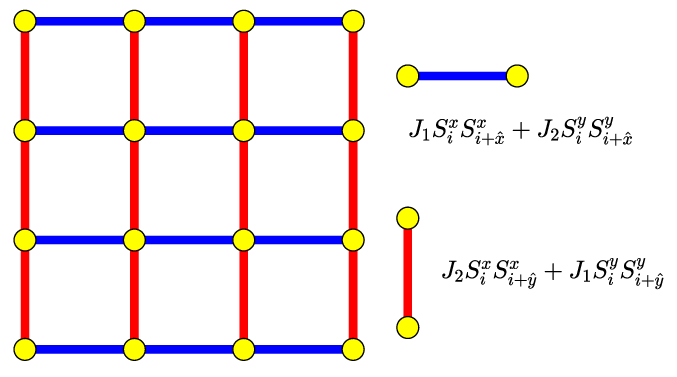}
\vskip 0.4cm
\caption{
	Illustration of the XY model with generic direction-dependent interactions, Eq.(\ref{H_ext_compass}). On the sites of the square lattice is a two-component classical XY spin.
	The bond-direction dependent interactions in this model are given by $J_1S^x_iS^x_{i+\hat{x}} + J_2 S^y_i S^y_{i+\hat{x}}$ in the $x$ direction and $J_2S^x_i S^x_{i+\hat{y}}+ J_1 S^y_i S^y_{i+\hat{y}}$ in the $y$ direction. This form of interaction possesses a symmetry operation that combines both internal and spatial ${\mathcal C}_4$ transformations, where neither the spatial operation nor the internal operation alone are symmetries of the system. }
\label{lattice}
\end{figure}

While the sub-system symmetries are fascinating as a theoretical problem, from the point of view of magnetism in Mott insulators, they don't appear naturally;  their presence in a model is a fine-tuned accident. Our goal thus is to find a simple extension of the model, Eq.~(\ref{H_compass}), which would model a generic magnetic Mott insulator, {\em i.e.} one that has ${\cal C}_4$ of the square lattice but without the extra subsystem symmetry. Here, we propose to study the model,
\begin{eqnarray}
      H_{\rm gCM}&=&- \sum_{i}  \left (J_1 S_i^x S_{i+\hat x}^x + J_2 S_i^y S_{i+\hat x}^y\right.\nonumber \\ 
      &+&\left. J_1 S_i^y S_{i+\hat y}^y+ J_2 S_i^x S_{i+\hat y}^x 
      \right ),
\label{H_ext_compass}
\end{eqnarray}
which for our purposes is a ``generic compass model" (gCM). The ${\vec S}_i=(S^x_i, S^y_i)$ again a two-component unit vector spin defined on lattice sites $i$ (Fig.\ref{lattice}). The first line describes spins interacting on $x$-oriented bonds, and the second on $y$-oriented bonds. Generically our model possesses the ${\cal C}_4$ symmetry of the compass model. In addition, the subsystem symmetries of the CM are now reduced to a global $\vec S\rightarrow -\vec S$ Ising symmetry.
We parameterize the exchange constants by setting $J_1=1, J_2=1-\alpha > 0$. We are thus left with two tuning parameters, $\alpha$ and the temperature $T$. An interesting feature of this generic compass model is that it has some well-known limiting cases; when $\alpha=0$, the model reduces to the 2D XY model, whereas when $\alpha=1$ it is the standard square lattice CM~\cite{compass_rev, Mishra}.

The goal of this paper is to establish, using both numerical and field-theoretic arguments, the phase diagram of this model. We will show that relaxing the fine-tuned symmetry makes a dramatic change to the phase diagram. The phase transitions, in particular, are very rich with critical phenomena that have continuously varying exponents analogous to the Ashkin-Teller (AT) model~ (for pedagogical reviews, see, e.g. ~\cite{Baxter, Nienhuis, Delfino}). We find, in addition, that the manner in which this phase diagram connects to the well-studied limiting cases of $\alpha=0$ and $1$ is  rather intriguing. 

The paper is organized as follows. In Sec.~\ref{sec:pd}, we give an overview of the main results of this paper--the phase diagram of the model Eq.~(\ref{H_ext_compass}) and its two regimes of transitions, the line of continuous critical points  and the first order regime. We then turn to the technical details: Sec.~\ref{sec:numeric} gives a summary of our Monte Carlo numerical method and the observables we study. Sec.~\ref{sec:cont} presents RG arguments in the small $\alpha$ regime and numerical evidence that the KT transition of the XY model at $\alpha=0$ evolves into a critical line with continuously varying exponents of the Ashkin-Teller type and Sec.~\ref{sec:first} presents evidence that the transition for larger $\alpha$ eventually becomes first order before meeting up with the compass model at $\alpha=1$. Finally, we provide a summary and outlook in Sec.~\ref{sec:sum}.

\begin{figure}[hbt!] 
\centering
\setlength{\abovecaptionskip}{20pt}
\includegraphics[angle=0,width=0.5\textwidth]{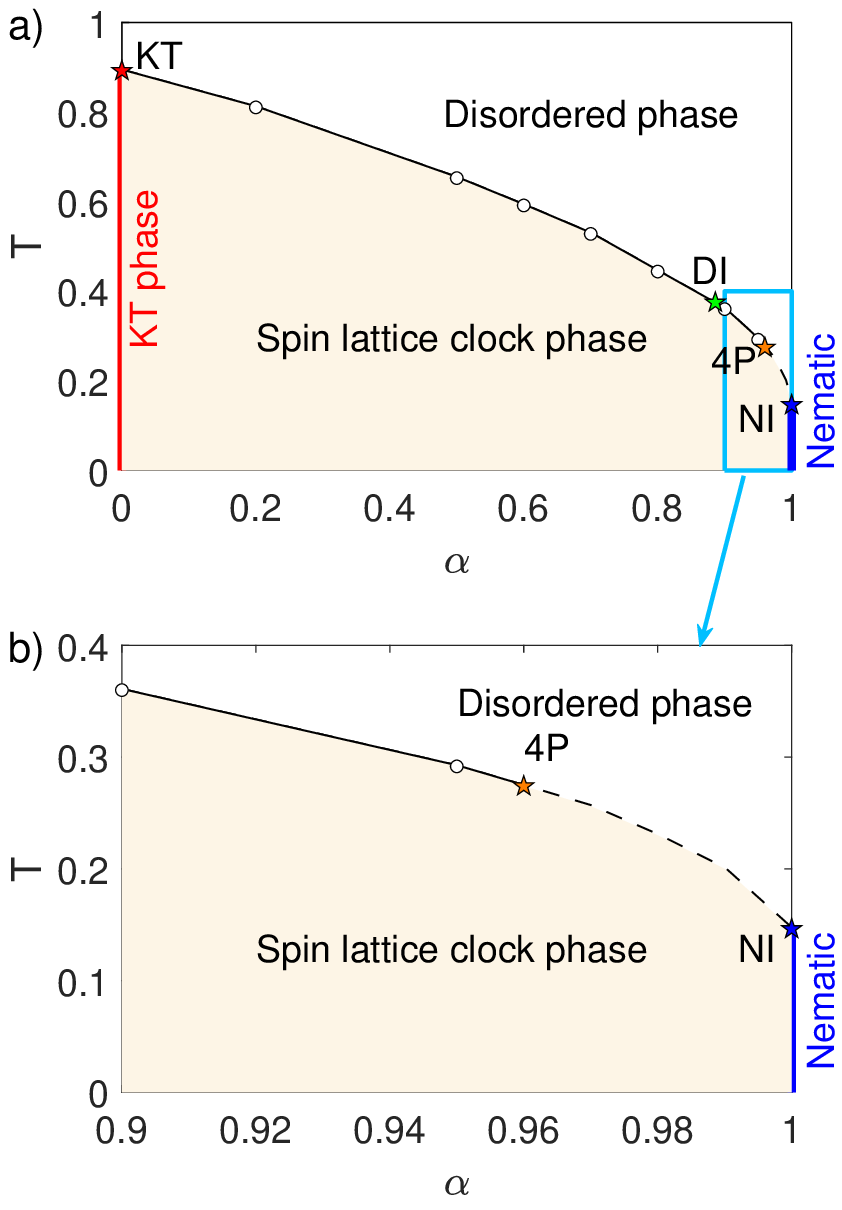}
\vskip 0.4cm
\caption{The main result of this work: the phase diagram of the generalized compass model, Eq.~(\ref{H_ext_compass}). In a), we show the phase diagram for our model. The critical region of the BKT phase is denoted by a solid red line, while the BKT phase transition point is indicated by a red star marked KT. The nematic phase of the compass model is represented by a solid blue line, with the nematic transition, which has been shown in previous works to be an Ising transition marked by a blue star (NI). The continuous order-disorder phase transition with continuously varying critical exponents of the AT type is depicted by a solid black line; the first-order phase transition is illustrated by a dashed black line. The decoupled Ising (DI) point is marked by a green star. b) is a closer examination of the phase diagram within the range $\alpha=0.9\sim 1$. The 4-state Potts point (4P) is marked by an orange star.}
\label{pd}
\end{figure}

\section{\label{sec:pd}Results: Phase Diagram}

\begin{figure}[!th]
\centering
\setlength{\abovecaptionskip}{20pt}
\includegraphics[angle=0,width=0.5\textwidth]{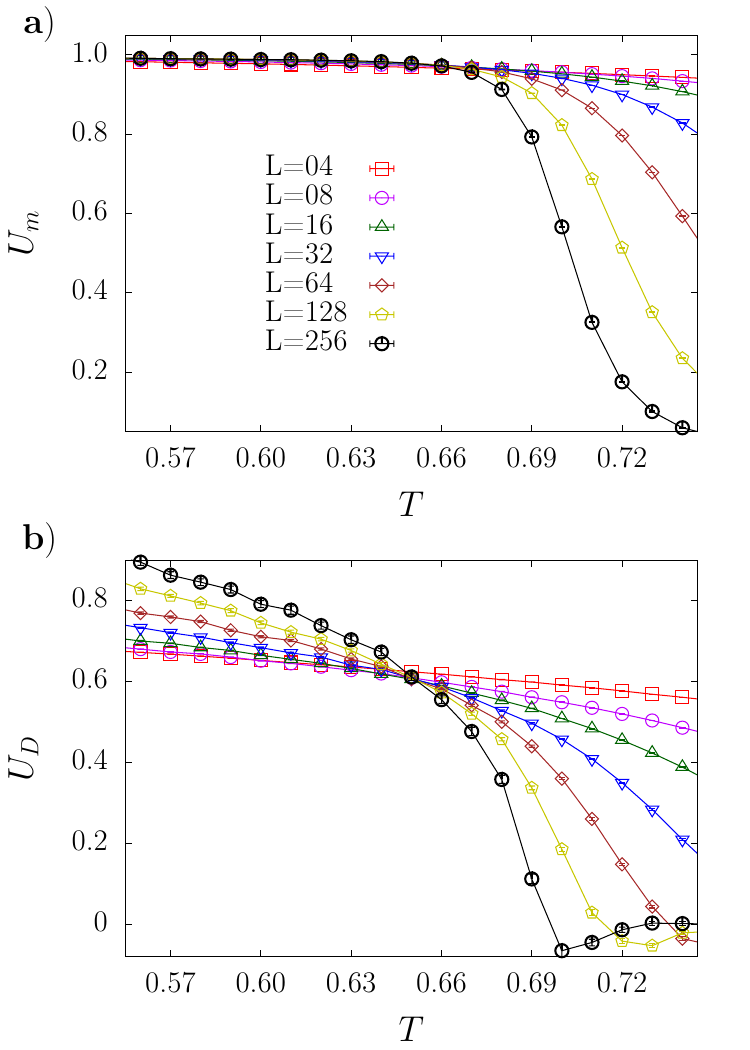}
\vskip 0.4cm
\caption{Binder cumulants as functions of temperature for a) magnetic order $U_m$ and b) nematic order $U_D$ for various system sizes $L$ at $\alpha=0.5$. 
While $U_D$ clearly exhibits a crossing point, showing a distinct signature of a continuous phase transition at $T_c\approx 0.649(5)$, it is observed that $U_m$ has a phase transition at the same $T_c$, though a clear crossing point is absent. We interpret this as arising from crossovers from the KT physics. We note that a clear crossing point does emerge for both $U_m$ and $U_D$ at larger $\alpha$, see Fig.~\ref{U0.9}. The analysis in Fig.~\ref{rg} presents alternate evidence for a direct transition in the magnetic order parameter at the same $T_c$ within errors at $\alpha=0.5$. }
\label{U0.5}
\end{figure}

Before turning to the details of how our conclusions were obtained, we discuss the main result of this work: the phase diagram of our model and the nature of the phase transitions, which are summarized in Fig.~\ref{pd} a) [ b) is a detail of a) in the region close to $\alpha=1$]. Our results are obtained using both Monte Carlo numerical simulations and some general field-theoretic arguments, as described in detail in the subsequent subsections. 

Let us start with the well-known limiting cases. When $\alpha=0$, the model becomes the 2D XY model, which goes through a Berezinskii-Kosterlitz-Thouless phase transition(BKT) \cite{berezinskii1972,kosterlitz1973,kosterlitz1974} at a finite temperature 
$T_C=0.8935(1)$\cite{BKT_Sandvik}. The BKT phase transition point is marked with a red star in the phase diagram. As is well known, it separates a high-$T$ phase with exponential spin-spin correlations from a low-$T$ phase with power-law correlations; the latter is indicated by the bold red line and labeled ``KT phase."  When $\alpha=1$, our model becomes the CM, which as a function of temperature goes from a high-$T$ disordered phase to a low-$T$ nematic ordered phase that breaks the ${\cal C}_4$ symmetry but preserves the $\vec S \rightarrow -\vec S$ symmetry. The transition takes place 
at a finite temperature $T_C=0.14621(2)$ \cite{Wenzel_2010} and has been demonstrated to be in the Ising universality class. The compass transition point is marked with the blue star in the phase diagram.

Our contribution is the phase diagram in the interior region, $\alpha \in (0, 1).$ First the vast region labeled ``spin-lattice clock phase" corresponds to a phase where both ${\cal C}_4$ and the $\vec S \rightarrow -\vec S$ symmetry are spontaneously broken. The phase transition from this ordered phase to the high-T disordered phase has two different regimes. For $0< \alpha \leq  \alpha_{\rm 4P}$  the transition is continuous but with continuously varying exponents, of the kind found in the AT model, in which the order parameter anomalous dimension $\eta=1/4$ throughout but the exponent $\nu$ varies from $\infty$ at the KT transition to $\nu=2/3$ at $\alpha_{\rm 4P}\approx 0.963$ when the transition is in the universality class of the four-state Potts model.  Interestingly, for a special value $\alpha_{\rm DI}\approx 0.885$, the critical point is described as a fixed point of two decoupled Ising models.  In contrast to the AT model or the four-state clock model, where such a fixed point is reached because the models microscopically become two decoupled Ising models, in our model, there is no microscopic decoupling and we find that the decoupling is an emergent phenomenon that results from renormalization group flow. Finally, in the interval $\alpha_{\rm 4P}< \alpha < 1 $, we find the transition is first order. The first-order transition must become very weak for both $\alpha_{\rm 4P}$ and $1$ to match the continuous transitions at these couplings. Our numerical data uncovers an unexpected aspect of this phase diagram, that the order-disorder transition (at which both ${\cal C}_4$ and the $\vec S \rightarrow -\vec S$ break spontaneously) merges with the Ising-nematic transition (at which ${\cal C}_4$ is broken but $\vec S \rightarrow -\vec S$ is part of a quasi-local symmetry and hence cannot break) in the compass model. Naively, one might have expected instead that the transition split into two Ising transitions with an intermediate nematic phase, but we do not find this in our numerical study.

\section{\label{sec:numeric}Numerical Method}

We perform Monte Carlo simulations using hybrid updates \cite{Wenzel_2010} with both Wolff clusters \cite{Wolff} and the ordinary Metropolis method. 
Several observables are used to describe the properties of the phases and critical behaviors.


The magnetic order parameter is defined as
\begin{equation}
\langle m^2 \rangle =\langle m_x^2 \rangle +\langle m_y^2 \rangle,
\end{equation}
where
\begin{equation}
m_x^2=\left(\frac 1N \sum_i S_i^x \right)^2,\mbox{~~~} m_y^2=\left(\frac 1N \sum_i S_i^y \right)^2.
\end{equation}
The Binder cumulant of the magnetic order parameter is then obtained 
\begin{equation}
U_m=2-\frac{\langle m^4 \rangle }{\langle m^2 \rangle ^2}.
\end{equation}

Following Wenzel et al. \cite{Wenzel_2010}, we also define the energy difference order parameter describing the nematic ordered phase
\begin{equation}
\langle D^2 \rangle =\langle (E_x-E_y)^2 \rangle, 
\end{equation}
where
\begin{equation}
E_x=\frac 1N \sum_i S_i^x S_{i+\hat x}^x, \mbox{~~~} E_y=\frac 1N \sum_i S_i^y S_{i+\hat y}^y
\end{equation}
are the $x$-component energy along the $x$ direction and the $y$-component energy along the $y$ direction, respectively.

We define a Binder cumulant based on $D$
\begin{equation}
U_D=\frac 12 \left(3-\frac{\langle D^4 \rangle }{\langle D^2 \rangle ^2}\right).
\end{equation}

The spin-spin correlation is defined as
\begin{equation}
C_S(\vec r)=\langle \frac 1N \sum_i \vec S_i \cdot \vec S_{i+\vec r} \rangle.
\end{equation}
Similarly, we define the correlation of the nematic order parameter:
\begin{equation}
C_D(\vec r)=\langle \frac 1N \sum_i D_i D_{i+\vec r} \rangle.
\end{equation}
with $D_i=S_i^x S_{i+\hat x}^x-S_i^y S_{i+\hat y}^y$.
We often present the maximum distance correlation which is defined as,
\begin{equation}
C_S(L/2)=C_S(\frac L2 \hat x + \frac L2 \hat y),
\end{equation}
and
\begin{equation}
C_D(L/2)=C_D(\frac L2 \hat x + \frac L2 \hat y).
\end{equation}

\section{\label{sec:cont}Continuous Transition: $0< \alpha \leq \alpha_{4P}$}

\begin{figure}[tbh!]
\centering
\setlength{\abovecaptionskip}{20pt}
\includegraphics[angle=0,width=0.5\textwidth]{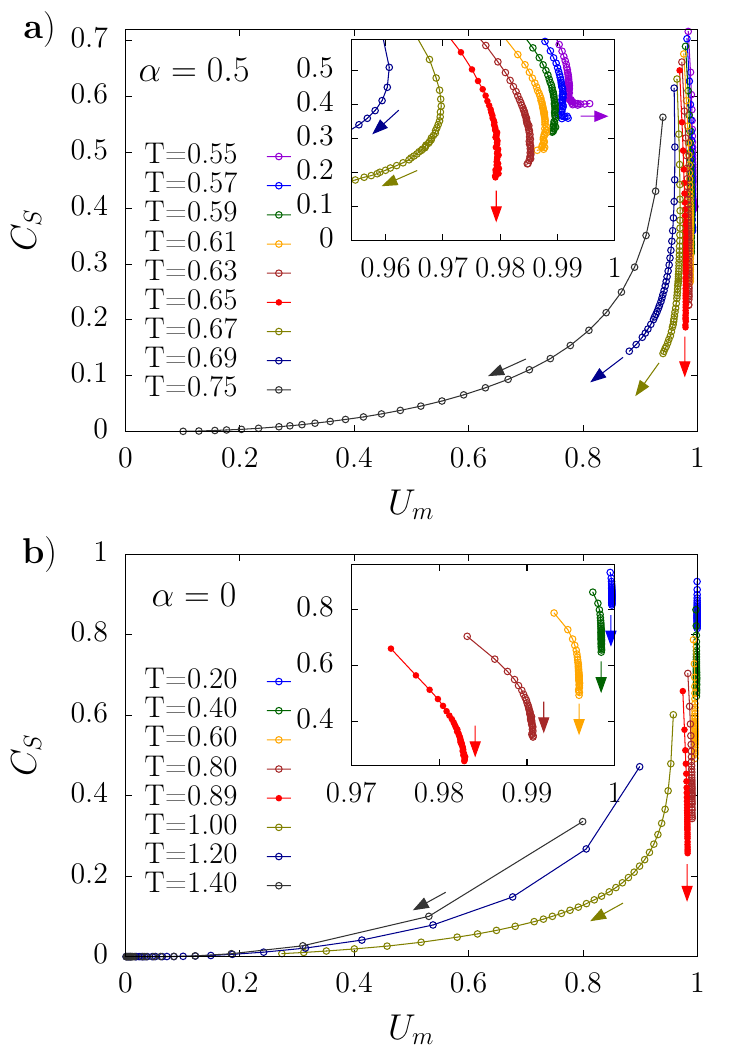}
\vskip 0.4cm
	\caption{Finite-size scaling data showing evidence for a direct magnetic order-disorder transition. The data is presented as a flow diagram at a) $\alpha=0.5$ and b) $\alpha=0$ (2D XY Model), with $x$-axis representing the Binder cumulant $U_m$ and the $y$-axis representing the spin correlation $C_s$. Each set of connected circles represents a fixed $T$ and sizes $L=2, 3, 4, \dots$, showing a numerical RG flow to fixed point values in the thermodynamic limit.
	a) The flow at $T=T_c \approx 0.650$ is denoted by red solid circles. The maximum lattice size is $L_{\rm max}=704$ for temperatures ranging from $T=0.61$ to $0.65$, 
	while for other temperatures, $L_{\rm max}=448$ or smaller. The flows at temperatures lower than $T_c$ converge towards locations where 
	$U_m=1$ and $C_S$ takes on a finite value. Conversely, flows at temperatures higher than $T_c$ converge towards locations where $U_m=0$ and $C_S=0$. This 
	behavior indicates the presence of an order-disorder phase transition. The arrows indicate the direction of increasing $L$. b) illustrates the numerical RG flow diagram for the 2D XY model ($\alpha=0$). The flow at $T=T_{KT}$ is 
	also marked with red solid circles. The maximum lattice size is $L_{\rm max}=224$. Flows at temperatures lower than $T_{KT}$ converge 
	towards locations with finite $U_m$, while $C_S=0$, consistent with the power law KT phase.}
\label{rg}
\end{figure}

\subsection{Field Theory for $\alpha \ll 1$}

 We start our analysis by studying the $\alpha \ll 1$ limit, in which we perturb around the XY model. The effective action of the XY model is well known to take the sine-Gordon form $S_{\rm SG}= \int d^2 x \left [\frac{K}{2}(\vec \nabla \theta) ^2 + \lambda_v \cos(2\pi \phi)\right ]$, where $K$ is the stiffness and $\lambda_v$ the vortex fugacity. The RG flows of this theory are now widely accepted to have a low-$T$ (large-$K$) region in which $\lambda_v$ is irrelevant, sometimes called the ``KT phase", and a high-$T$ phase at which $\lambda_v$ is relevant separated by a critical coupling, $K_c=2/\pi$. While the high-$T$ phase with exponential decaying correlations is generically expected to be stable for small $\alpha$, the fate of the low-$T$ power-law phase needs investigation. We can frame this as an RG picture by asking whether the leading perturbation introduced by the anisotropy of the $H_{\rm gCM}$ is relevant or irrelevant at the XY fixed point, which is itself described by the spin-wave action $S_0= \frac{K}{2}\int d^2 x (\vec \nabla \theta) ^2$ with $K>K_c=2/\pi$. Expanding the lattice model for $\alpha \ll 1$, we find that the leading anisotropic perturbation is,
\begin{equation}
\label{lambdaa}
S_a = \lambda_a \int d^2 x [(\nabla_x \theta)^2 -(\nabla_y \theta)^2] \cos(2 \theta). 
\end{equation}
As expected, the term changes sign under either $\pi/2$ rotations of space or spin but is invariant under a combination of the two, exactly as is expected for a lattice model with ${\cal C}_4$. Simple power counting at the spin wave fixed point established that $[\lambda_a] =-1/(\pi K)$ is  irrelevant throughout the low-$T$ phase, so it cannot by itself destabilize the power law phase. On the other hand, in the usual Wilsonian RG, $\lambda_a$ can generate the more symmetric term in RG flow,
\begin{equation}
S_4 = \lambda_4 \int d^2 x \cos(4 \theta) 
\end{equation}
This term is the continuum version of the fourfold magnetic field, which is exactly marginal at the KT phase transition.  Since $S_4$ is generated by $S_a$ and is much more relevant, we make the reasonable assumption that we can neglect $S_a$ and the effective action for the criticality is simply given by $S_{\rm SG}+S_4$. The $S_{\rm SG}+S_4$ theory has been analyzed extensively, $S_4$ completely destroys the power law phase, replacing it with a fourfold ``clock"\ phase. The RG flows for this theory for small $\lambda_4$ feature a line of critical points with continuously varying exponents of the AT type that separate the clock phase from the high-temperature disordered phase\cite{Jose-clock}. In our model, because of the spin-lattice locking captured by $S_a$, the breaking of spin symmetry triggered by $S_4$ also breaks the lattice symmetry. In this way, $S_a$ does not affect the critical behavior, but it does affect the details of the ordered phase.

We provide evidence supporting the hypothesis of the above RG picture for our lattice model $H_{\rm gCM}$ in two steps; first, we show that at $\alpha \neq 0$ the KT power-law phase is destroyed, and there is a direct order-disorder transition. This demonstrates in our lattice model that $\lambda_a$, although formally irrelevant in the KT phase, generates couplings that are relevant: even though there is no long-range order in the KT phase, this turns on immediately for $\alpha \ll 1$ (as shown in Fig.~\ref{pd}); then in the second step we study the nature of the transition in detail and present various pieces of evidence for the AT criticality, including continuously varying exponents of the correct form predicted by the AT theory. 

We start with the first step, where we show numerically that the power-law phase gives way to a ``spin-lattice clock phase" once $\alpha \neq 0$, as shown in our phase diagram Fig.~\ref{pd}.
Figure \ref{U0.5} shows the finite-size behaviors of the Binder cumulants $U_m$ and $U_D$ at $\alpha=0.5$. Interestingly, while $U_m$  exhibits peculiar characteristics (we attribute this to crossovers that arise from the proximity to the power-law KT phase, which is riddled with notoriously complicated finite-size corrections; this limits our ability to study smaller $\alpha$), with its value 
appears to be very close to one across different system sizes at low temperatures, 
 $U_D$ demonstrates clearly crossing points, showing typical behavior associated with a continuous phase transition.

In order to further elucidate the absence of a KT power-law phase at finite-$\alpha$, we have employed a numerical flow-diagram analysis of the flow of our model, as depicted in 
Fig.~\ref{rg}. The $x$-axis represents the Binder cumulant $U_m$, while the $y$-axis represents the spin correlation $C_S$. For our model, in the disordered phase (high temperature), the flows converge towards the disordered fixed point with $U_m=0$ and $C_S=0$ as the system size $L$ tends to infinity. In the ordered phase (low temperature), the flows converge towards the magnetically ordered fixed point with $U_m=1$ and $C_S>0$ as $L$ approaches infinity. At the 
critical point ($T_C \approx 0.650$), the flow converges towards the nontrivial fixed point with $U_m \approx 0.98$, while $C_S(L/2)=0$ as $L$ tends to infinity. In contrast, for the 2D XY model (shown in the lower panel of Fig.~\ref{rg}), in the disordered phase (high temperature), the flows also converge towards the disordered fixed point with $U_m=0$ and $C_S=0$ as $L$ tends to infinity. However, in the KT critical region (low temperature), the flows converge towards different points on the $x$-axis, with $U_m$ taking on a finite value while $C_S=0$, 
as expected in the power law phase. While we expect that the same behavior is valid at arbitrary small $\alpha$ based on our RG argument, we cannot reach large enough lattices for $\alpha<0.5$ to demonstrate this convincingly numerically because this regime is dominated by crossover behavior, which is expected from the fact that the RG generated $\lambda_4$ is expected to be very small, so it requires very large lattices to observe its relevance. Interestingly, the behavior of the model for $\alpha \neq 0$ differs from the 2D XY model, which has a critical phase when $T<T_{KT}$. It is also different from the 2D $90^\circ$ compass model, which has a nematic-ordered low-temperature phase but not a magnetically ordered phase\cite{Mishra}\cite{Wenzel_2010}. Indeed, the spin-lattice clock phase is a unique phase that has both spin and lattice rotation symmetry breaking.

We now turn to the details of our analysis of the phase transition that separates the spin-lattice clock phase and the high-$T$ disordered phase, which we conclude is of the AT type with continuously varying exponents. Since this analysis has many different aspects, we have broken it up into subsections.

\subsection{Locating transition points}

The first step in studying the critical behavior is to locate the phase transition. According to standard finite-size scaling theory\cite{Nightingale, Fisher_Baber}, the Binder cumulant
$U_m(L)$ converging to 1 with increasing system size indicates the existence of magnetic order while tending to zero with increasing system size implies that the system is in the magnetic 
disordered phase. The crossings of curves for different sizes indicate a critical point separating the two phases.

We adopt the standard $(L,2L)$ crossing analysis for the Binder cumulant to estimate the 
critical point and critical properties; see, e.g., the Supplemental Material of \cite{Shao_2016}. 
The crossings point $T_c(L)$ of the Binder cumulant (both $U_m$ and $U_D$) $U(T, L)$ and $U(T, 2L)$ is expected to converge to the critical point $T_c$ in the following way
\begin{equation}
   T_c(L)=T_c +\sum_i a_i L^{-1/\nu-\omega_i}
   \label{crossp}
\end{equation}
where $\nu$ is the correlation length exponent, $\omega_i>0$ are irrelevant exponents. $a_i$ are  unknown coefficients. 

\begin{figure}[t]
\centering
\setlength{\abovecaptionskip}{20pt}
\includegraphics[angle=0,width=0.5\textwidth]{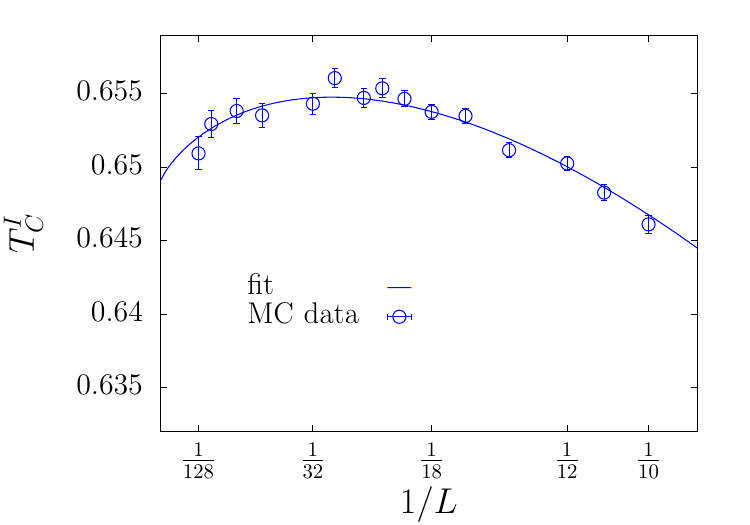}
\vskip 0.4cm
\caption{
	Crossing data analysis of $U_D$ to obtain $T_C^I$ for $\alpha=0.5$. 
	Since the crossing points first increase, then decrease with $L$, the scaling form Eq. (\ref{crossp}) with two powers are used and $T_C^I=0.649(5)$ is estimated,
	with the fit windows $L=10\sim 128$. 
	}
\label{cp0.5}
\end{figure}

As discussed for $\alpha=0.5$, finding the crossing points for $U_m$ (shown in Fig. \ref{U0.5}) proves to be challenging. Since the crossing of $U_D$ is more well-defined, we have employed the standard $(L,2L)$ crossing analysis for $U_D$ and determined $T_C=0.649(5)$. The data of the crossing points $T_c(L)$ as a function of the inverse size $1/L$, along with the fitting line, are presented in Fig. \ref{cp0.5}. 
Since the crossing points first increase and then decrease with $L$, we 
have to use two irrelevant exponents in Eq. (\ref{crossp}) to fit the data\cite{MaNvsen}. 
The fitting window used is $L=10\sim 128$, yielding a reduced $\chi^2=1.03$.

We apply similar analyses for different $\alpha$ and obtain critical points $T_C^I$, as listed in Table~\ref{cp}.

\begin{figure}[t]
\centering
\setlength{\abovecaptionskip}{20pt}
\includegraphics[angle=0,width=0.5\textwidth]{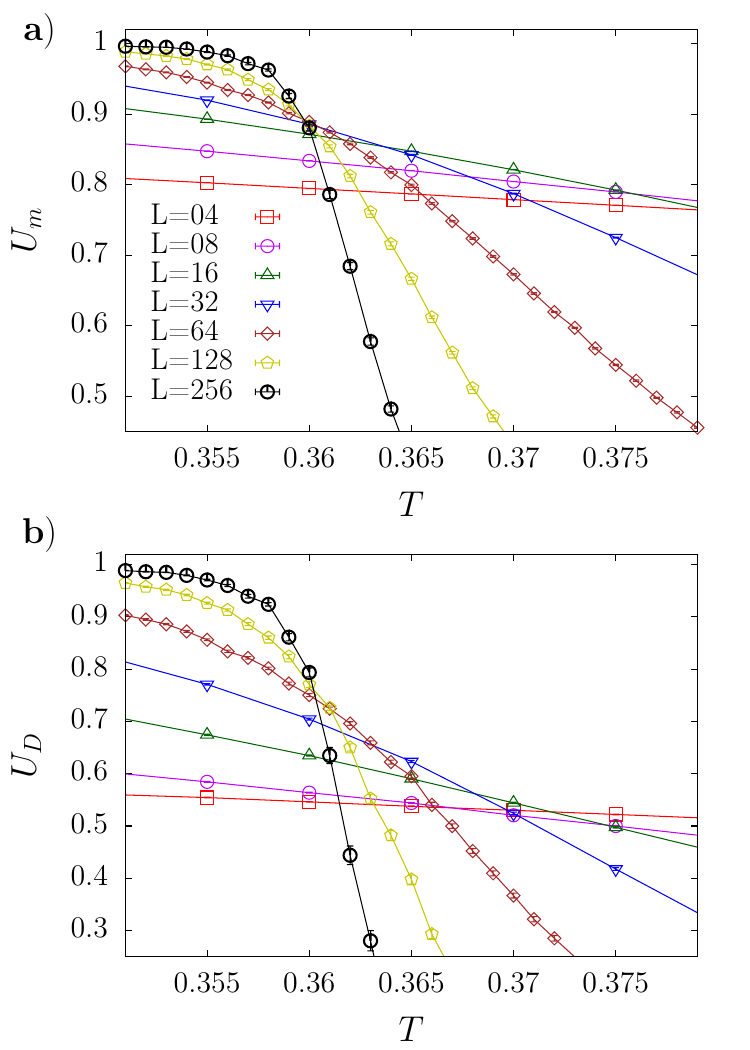}
\vskip 0.4cm
	\caption{Binder cumulants a)$U_m$ and b) $U_D$ as functions of temperature $T$ at $\alpha=0.9$. 
Both $U_m$ and $U_D$ exhibit distinct crossing points for different sizes $L$, indicating the presence of a continuous phase transition. }
\label{U0.9}
\end{figure}

The crossing points of $U_m$ for different sizes $L$ become evident as the parameter $\alpha$ increases. This phenomenon can be observed in both $U_m$ and $U_D$ when $\alpha$ reaches a sufficiently large value, as shown in Fig. \ref{U0.9}, which displays the Binder cumulant for $\alpha=0.9$. 
As a result, the conventional $(L,2L)$ crossing analysis can also be applied to $U_m$ when $\alpha$ is sufficiently large. The critical points obtained are listed in Table \ref{cp} as $T_C^{II}$.

\begin{table*}[htbp]
\begin{ruledtabular}
  \centering
  \caption{ Critical temperatures $T_C$ and exponents $\eta$ and $\nu$ for various $\alpha$ obtained using different methods.
	  $T_C^I$ is the critical temperature obtained by performing a crossing analysis of the Binder cumulant $U_D$ 
	  of pairs of sizes $L$ and $2L$. 
	  $T_C^{II}$ is the critical temperature determined using the crossing points of the curves for $U_m$. 
	  We first got $\eta({T_c})$, $\eta(T_c^{(-)})$, and $\eta(T_c^{(+)})$, which are anomalous exponent $\eta$ obtained by calculating 
	  spin correlation  at $T_c^I$, $T_c^I-\sigma$, and $T_c^I+\sigma$,
	  respectively, with $\sigma$ the statistical error of $T_c^I$. Then conclude $\eta$.
	  $T_C^{III}$ and $1/\nu$ are obtained by 
	  fitting the data collapse of $U_D$  
   (except for $\alpha=0.5$ where we use $m^2L^{1/4}$, see Tab. \ref{tab:nu} for details). $\eta_D$ is the anomalous dimension of the nematic order parameter $D$. }
    \begin{tabular}{ccccccc}
    $\alpha$ & $T_c^I$ & $T_c^{II}$ & $T_c^{III}$ & $\eta$ & $1/\nu$ & $\eta_D$ \\
    \hline
    0 & & & 0.8935(1)\cite{BKT_Sandvik} &   & 0 & \\
    0.2 & & & 0.8104(2) & & 0.23(4) & 0.89(2) \\
    0.5 & 0.649(5) & & 0.649(2) & 0.245(13) & 0.37(2) & 0.85(3) \\
    0.6   & 0.594(3) & & 0.5975(4) & 0.255(10) & 0.49(3) & 0.79(4) \\
    0.7   & 0.5292(6) & & 0.531(5) & 0.253(4) & 0.60(3) & 0.75(2) \\
    0.8   & 0.4546(7) & & 0.4555(1) & 0.245(15) & 0.75(6) & 0.642(7) \\
    0.9   & 0.3596(3) & 0.35957(6) & 0.36019(7) & 0.225(25) & 1.07(2) & 0.48(2) \\
    0.95   & 0.29175(5) & 0.29173(3) & 0.29185(2) & 0.22(3) &  1.38(2) & 0.32(2)\\
    1 & & & 0.14621(2)\cite{Wenzel_2010} &  &  & 1 \\
    \end{tabular}%
  \label{cp}%
\end{ruledtabular}
\end{table*}%

\subsection{Critical Exponents: Evidence for Ashkin-Teller criticality}

We now present evidence for Ashkin-Teller criticality along the line of critical points by studying three critical exponents $\eta,\nu$ and $\eta_D$, the anomalous dimension of $\vec S$, the correlation length exponent and the anomalous dimension of the nematic order parameter $D$. The most striking feature of the criticality is the existence of continuously varying critical exponents as one moves along the critical line, which are all controlled by a single parameter $g_R$, the coupling constant in a Coulomb gas description. An exception is the critical exponent associated with the Ising field of the Ashkin-Teller model $\eta=1/4$,  which is constant along the line and independent of $g_R$. We shall verify these features in our numerical study of the model $H_{\rm gCM}$.

\subsubsection{$\eta$}

\begin{figure}[t]
\centering
\setlength{\abovecaptionskip}{20pt}
\includegraphics[angle=0,width=0.5\textwidth]{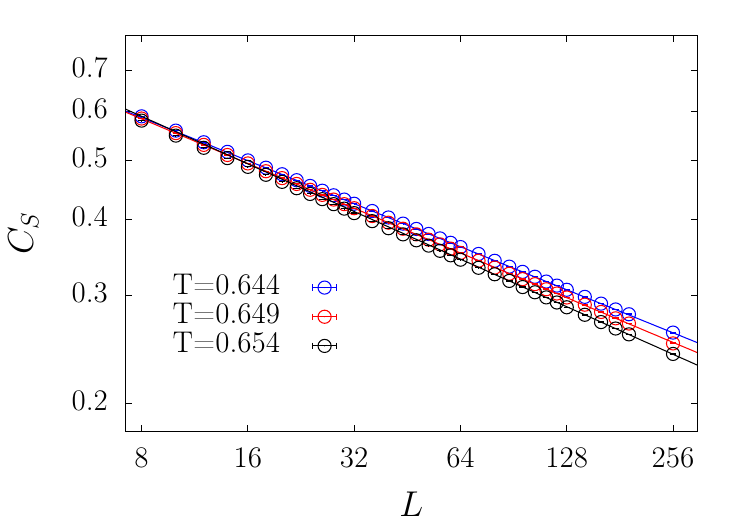}
\vskip 0.4cm
\caption{ Estimate anomalous exponent $\eta$ using spin correlation.
	The longest distance spin correlation $C_S(L/2)$ at $T_C^{(-)}=0.644, T_C^I=0.649, T_C^{(+)}=0.654$ of $\alpha=0.5$ are shown on a log-log scale.
	The Monte Carlo simulation data are indicated by open circles, and the straight lines are the power law fits to the data. 
	The power law fits are statistically sound for $T=0.649$, with $\eta=0.2427(2)$ and a reduced $\chi^2$ value of 1.06, over the fit window $L=24\sim 256$;
	for $T=0.644$, with $\eta=0.2336(2)$ and a reduced $\chi^2=0.955$, over the fit window of $L=24\sim 256$; for $T=0.654$
	with $\eta=0.258(1)$ and a reduced $\chi^2$ value of 1.04, over the fit window $L=80\sim 256$. }
\label{cor0.5}
\end{figure}

As discussed above, in the AT universality class, the exponent $\eta=1/4$ is a constant throughout. We now present our numerical evidence for this behavior in our model. 
At the estimated critical points, we calculate the spin correlation $C_S(L/2)$. 
From the decay of $C_S(L/2)$ we can determine the anomalous scaling dimension $\eta$ according to the following
finite-size scaling formula $C_s(L/2)\propto L^{-\eta}$. Since the estimated $T_c$ has statistical errors, we calculate $C_S(L/2)$ at the upper bound, center value, and the lower bound 
of the estimated $T_c$ for each $\alpha$. The powers $\eta$ found by fitting the power law to the data at the three values thus lead to a reasonable estimate
of $\eta$ and its error bars. For example, for $\alpha=0.5$, we calculate $C_S(L/2)$ for several system sizes $L$ at $T=0.644$, $T=0.649$ and $T=0.654$, as shown
in Fig. \ref{cor0.5}. The circles are Monte Carlo data. the lines are fitted functions.  We then
obtain $\eta=0.2336(2)$ at $T=0.644$, $\eta=0.2427(2)$ at $T=0.649$ and $\eta=0.258(1)$ at $T=0.654$ from fits to the data. 

We have done similar analyses for other $\alpha$. The estimates of $\eta$ and the errors are listed in Table~ \ref{cp}. From this Table,  we conclude that along the transition line, $\eta\approx 1/4$ does not change, and thus, the spin operator should be identified with the Ising field of the AT model.

\subsubsection{$\nu$}

In contrast to the constancy of the $\eta$ of the spin field, the exponent $\nu$ varies continuously in the AT model. We present an analysis of the behavior of this exponent in our model. 

\begin{table}[htbp]
\begin{ruledtabular}
  \centering
  \caption{A comparison of $1/\nu$ fit from $U_D$ and $m^2L^{1/4}$}
    \begin{tabular}{ccc}
    $\alpha$ & $1/\nu(U_D)$ & $1/\nu(m^2 L^{1/4})$ \\
    \hline
    0.2   & 0.23(4) & 0.22(6) \\
    0.5   & 0.38(15) & 0.37(2) \\
    0.6   & 0.49(3) & 0.43(3) \\
    0.7   & 0.60(3) & 0.57(2) \\
    0.8   & 0.75(6) & 0.71(2) \\
    0.9   & 1.07(2) & 1.04(2) \\
    0.95  & 1.38(2) & 1.40(3) \\
    \end{tabular}%
  \label{tab:nu}%
\end{ruledtabular}
\end{table}%

\begin{figure}[t]
\centering
\setlength{\abovecaptionskip}{20pt}
\includegraphics[angle=0,width=0.5\textwidth]{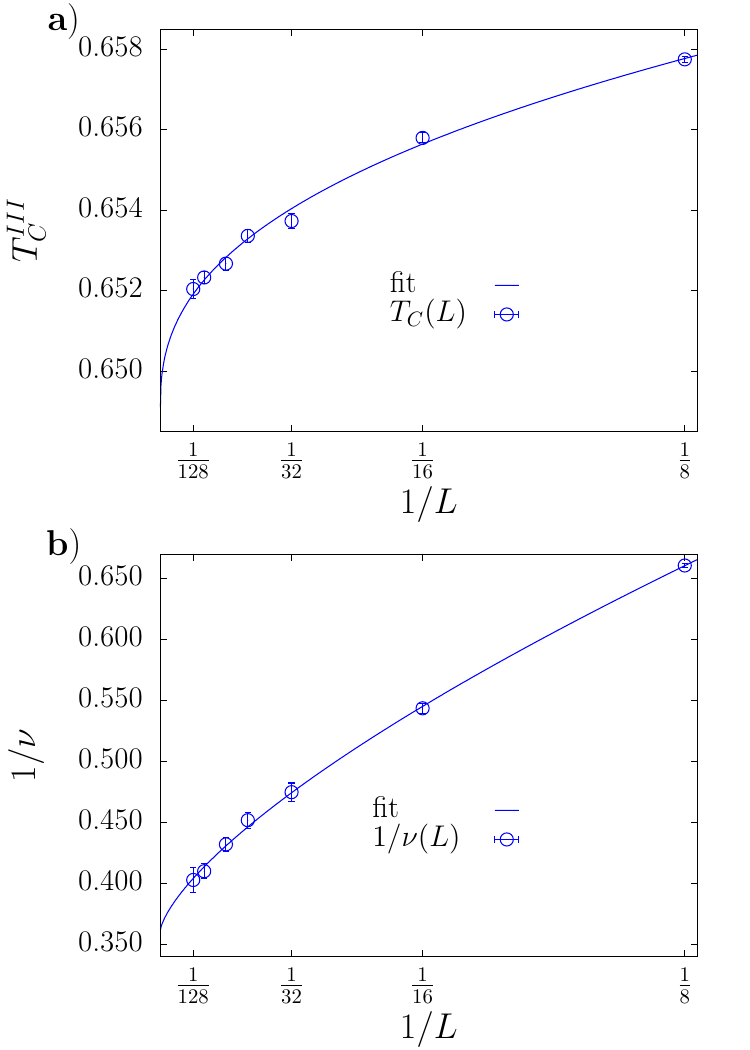}
\vskip 0.4cm
	\caption{The finite-size analysis of $T_C(L)$ and $1/\nu(L)$ to estimate $T_C^{III}$ and  $1/\nu$ at $\alpha=0.5$. 
	a) shows $T_C(L)$ and b) shows $1/\nu(L)$ obtained by polynomial fit 
 of the scaling function $f_1(tL^{1/\nu})$
 up to $P=3$ for pairs of sizes $L$ and $2L$. The solid lines are power-law fits 
 to $T_C(L)$ and 
 to $1/\nu(L)$, respectively.}
\label{dcl0.5}
\end{figure}

According to finite-size scaling theory, $U_D(T,L)$, $m^2(T, L) L^\eta$ is expected to behave in the standard way,
$U_D(t,L)=f_1(tL^{1/\nu})$ and 
$m^2(T, L) L^\eta=f_2(tL^{1/\nu})$, where $t=(T-T_c)/T_c$ is the reduced temperature. 
$f_1(x)$ and $f_2(x)$ are two scaling functions  and can be expanded to polynomials near the phase transition points, $f_{i}=\sum_{P=0} a^{(i)}_P x^{P}$
with $i=1, 2$ denoting the two scaling functions, respectively. Thus we can simultaneously obtain $T_C$ and $\nu$ by using polynomial fit for the scaling functions. Instead of fitting all our data to this asymptotic form, we attempt to take into account finite size corrections by fitting the scaling functions using data from different pairs of systems $(L,2L)$ in order to obtain $T_C(L)$ and $\nu(L)$. We then obtain our best estimates for $T_C$ and $\nu$ through extrapolation using the following forms, 
	$T_C=T_C(L)+a_1/L^{b_1}$,
	and 
	$1/\nu=1/\nu(L)+a_2/L^{b_2}$.
Figure \ref{dcl0.5} illustrates a finite-size analysis of $T_C(L)$ and $1/\nu(L)$,
obtained by polynomial fits of the scaling function $m^2L^\eta$ under the assumption that $\eta=1/4$,
for $\alpha=0.5$.  This method can be applied to different values of $\alpha$. The results for $1/\nu$ from both quantities are collected in Table~\ref{tab:nu}.
Finally, we present the data from $U_D$ in Table \ref{cp} as $T_C^{III}$ and $1/\nu$ (except for $\alpha=0.5$ for which we used the $m^2L^\eta$ estimates since they have smaller errors). 

\subsubsection{$\eta_D$}

 We now turn to the scaling of the nematic order parameter $D$ and its anomalous dimension, $\eta_D$. We have seen that the anomalous dimension of the spin field $\vec{S}$ of the gCM remains constant at $\eta=1/4$ and can be identified with the scaling dimension $x_h$ of the individual Ising variables of the Ashkin-Teller model. We identify the nematic order parameter $D$ with the so-called ``polarisation" operator (in the usual representation of the AT model as two coupled Ising models, this corresponds to the product of the two Ising variables) with scaling dimension $x_p$, which, like the thermal scaling dimension $x_t$, varies along the line of transitions. While $x_p$ and $x_t$ are distinct, they both are determined by the same Coulomb gas parameter $g_R$\cite{Nienhuis}. These operators have the scaling dimensions,
\begin{equation}
\begin{split}
&x_t=\frac{2}{g_R},\mbox{~~~} x_h=\frac{1}{8},\mbox{~~~} x_p=\frac{1}{2g_R}
\end{split}
\end{equation}

\begin{figure}[t]
\centering
\setlength{\abovecaptionskip}{20pt}
\includegraphics[angle=0,width=0.5\textwidth]{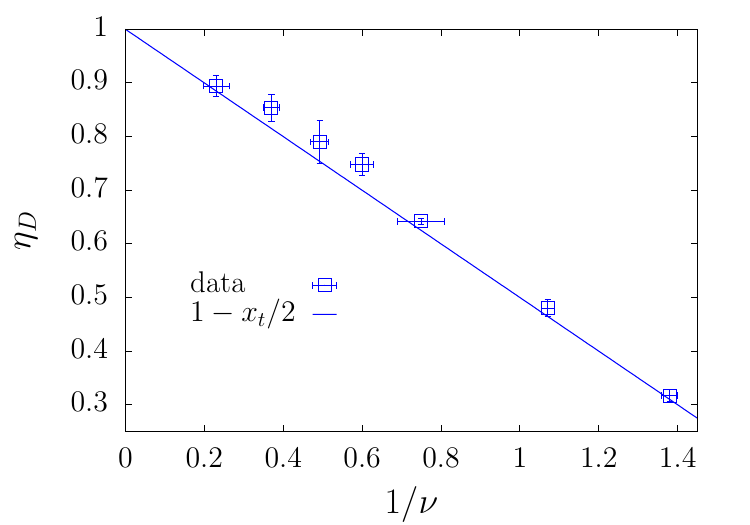}
\vskip 0.4cm
\caption{Comparison of the dependence of $\eta_D$ on $1/\nu$ from  Monte Carlo studies on the gCM model along the line of continuous phase transitions (shown as data with error bars), with the behavior expected from the Coulomb gas description of Ashkin-Teller criticality $\eta_D=1-\frac{1}{2\nu}$ (solid line). The good agreement establishes the Ashkin-Teller criticality in our gCM model  and the correct identification of the microscopic spin and nematic order parameters.}
 
\label{etaD}
\end{figure}

These scaling dimensions can be straightforwardly related to three exponents in our numerical simulations of the gCM. The correlation length exponent, $1/\nu = 2-x_t=2(1-1/g_R)$,
the anomalous dimension of the spin field $\vec S$ is $\eta=2x_h=1/4$, and the anomalous dimension of the nematic order parameter $\eta_D = 2 x_p = 1/g_R$. Combining these relations, we find the relation $\eta_D = 1-\frac{1}{2\nu}$, which is a non-trivial test of the identification of the nematic order parameter with the polarization operator and the general Ashkin-Teller criticality picture in our model.
 By using the same method as the one used to obtain $\eta$, we can obtain $\eta_D$ through Monte Carlo data. The results and expected relations are shown in Fig.\ref{etaD}, which show a good agreement.

\subsection{The 4-state Potts point}

\begin{figure}[t]
\centering
\setlength{\abovecaptionskip}{20pt}
\includegraphics[angle=0,width=0.5\textwidth]{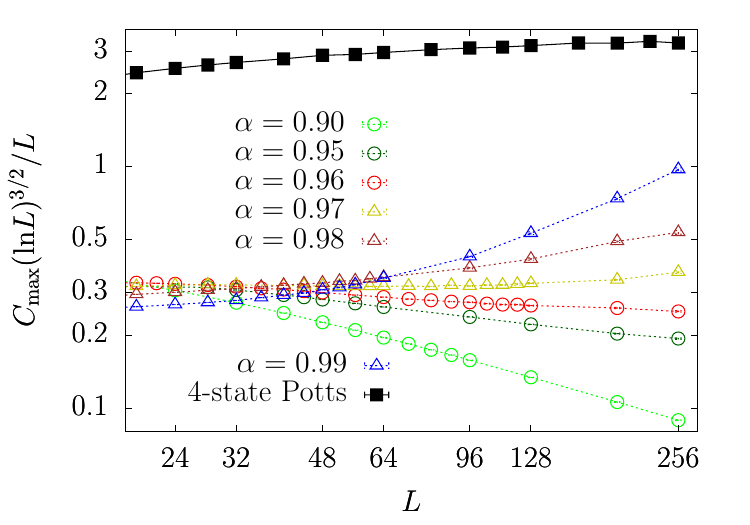}
\vskip 0.4cm
\caption{
Peak value $C_{\mathrm{max}}$ of the specific heat (suitably scaled)
for the gCM model for different $\alpha$ and a comparison of the same quantity for the 4-state Potts model. For Potts criticality it is expected that asymptotically, $C_{\rm max}=a L \ln(L)^{-3/2}$, where $a$ is a non-universal number. Therefore graphing $C_{\rm max}\ln(L)^{3/2}/L$ (as done above) should result in a constant value for large $L$. From our data it is apparent that the 4-state Potts point is observed  around $\alpha= 0.96\sim 0.97$.}
\label{maxcapchi}
\end{figure}

 We have presented detailed evidence for AT criticality described by a line of fixed points with $\nu$ continuously varying from $\infty$ at $\alpha=0$ to a decreasing finite value as $\alpha$ is increased. The behavior in the AT model was shown to arise from an effective mapping to a Gaussian model, which eventually becomes unstable at the 4-state Potts point ($\nu=2/3$) due to the emergence of another relevant operator~\cite{KADANOFF}.  We present various pieces of evidence that this happens around $\alpha = 0.96 \pm 0.01$. We first attempt to locate this 4-state Potts point in our model using the peak value $C_{\mathrm{max}}$ of the specific heat, which is well known to have characteristic logarithmic corrections at the critical point~\cite{Nauenberg, Salas, Cardy_Potts}.  
Figure \ref{maxcapchi} shows the finite size behavior of  $C_{\mathrm{max}}$. For reference, we have also shown the same quantity for the 4-state Potts model. 
 This analysis gives the results that the 4-state Potts point is around $\alpha=0.96\sim 0.97$.

\begin{figure}[t]
\centering
\setlength{\abovecaptionskip}{20pt}
\includegraphics[angle=0,width=0.5\textwidth]{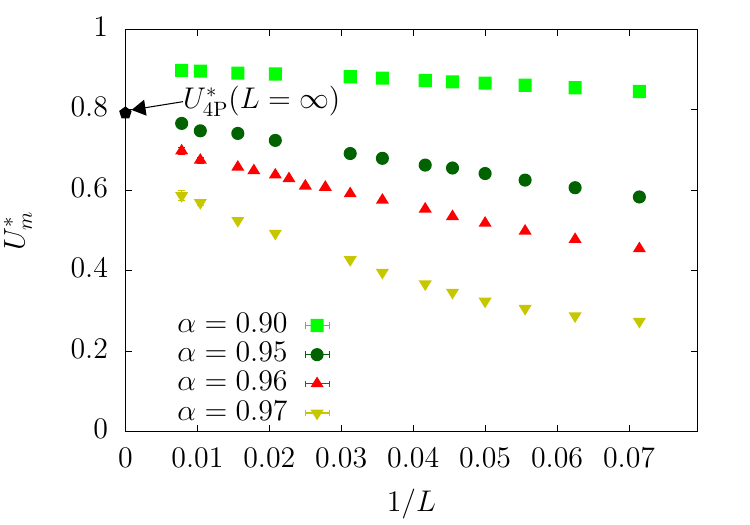}
\vskip 0.4cm
	\caption{Binder cumulant crossing points $U^*_m(L)$ comparison with the 4-state Potts point. This figure illustrates the crossing points $(L,2L)$ of the Binder cumulant $U^*_m(L)$ as the size $L$ increases. Markers of different colors represent data for various values of $\alpha$.  
 The black solid pentagon marker is the $U^*_{\mathrm{4P}}(L=\infty)=0.792(4)$ for the 4-state Potts model\cite{Jin,Jin2013}. While reliable extrapolations to the thermodynamic limit are difficult, it is clear that in the region $\alpha\sim 0.96 \pm 0.01$ our data extrapolates to values very close to the known four state Potts point, $U^*_{\mathrm{4P}}$, providing further evidence for the four state Potts point in this regime.}
\label{um0}
\end{figure}

\begin{figure}[t]
\centering
\setlength{\abovecaptionskip}{20pt}
\includegraphics[angle=0,width=0.5\textwidth]{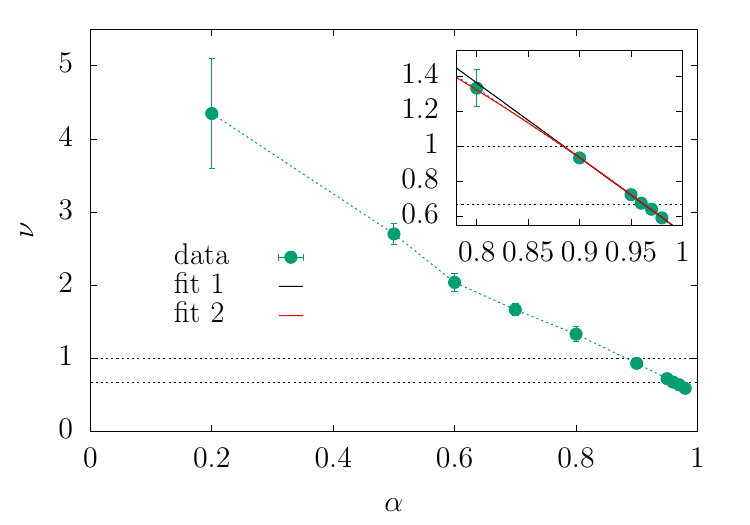}
\vskip 0.4cm
	\caption{ The dependence of $\nu$ on $\alpha$, and estimation of $\alpha_{\rm DI}$ and $\alpha_{\rm 4P}$. The green points represent $\nu$ from Table~\ref{cp}). The dashed black horizontal 
	lines indicate $\nu=2/3,1$ which are the universal values for the 4-state Potts model and Ising model. The data are fitted 
	using a linear function $f(x)=ax+b$ (shown as a black solid line) and a quadratic function $f(x)=ax^2+bx+c$ (shown as a red solid line) over 
	the range $\alpha=0.8\sim0.98$, which give consistent estimates. The linear fit gives $\alpha(\nu=1)=\alpha_{\rm DI}=0.885(23)$, which gives an estimate for the point at which $H_{\rm gCM}$ can be described as a pair of decoupled Ising fixed points (marked in Fig.~\ref{pd}). We can locate $\alpha_{\rm 4P}$ similarly,  $\alpha(\nu=2/3)=\alpha_{\rm 4P}=0.963(22)$, in general agreement with the estimates from $C_{\rm max}$ and $U^*_m$. }
\label{nu}
\end{figure}

Next, we analyze the Binder cumulant and the exponent $\nu$; since they both have universal values, they can identify the 4-state Potts point. Figure \ref{um0} shows the comparison of Binder cumulant $U_m$ of the 4-state Potts model, which gives the results that the 4-state Potts point is around $\alpha\approx 0.95$. The results for $\nu$ at different $\alpha$ are shown in Fig.\ref{nu}. From the data in Table \ref{cp}, we have obtained $\nu$ through data collapse using polynomial fitting. 
The data for $\alpha=0.2\sim 0.9$ are fitted using $U_D$, while the data for $\alpha=0.95\sim 0.98$ are obtained from $U_m$. 
A linear fit of the data yields 
the 4-state Potts point at $\alpha=0.963(22)$.

All of these indicators are consistent with our identification that $\alpha_{4P} = 0.96 \pm 0.01$.

\section{\label{sec:first}First order: $\alpha_{4P}<\alpha<1$}

Once we cross the 4-state Potts point, the line of transitions with continuously varying exponents must end due to the instability of the Gaussian fixed point theory. Beyond this, the nature of the phase transition changes;  we present numerical evidence now that there is still a direct order-disorder transition, but it becomes first order. 

\begin{figure}[t]
\centering
\setlength{\abovecaptionskip}{20pt}
\includegraphics[angle=0,width=0.5\textwidth]{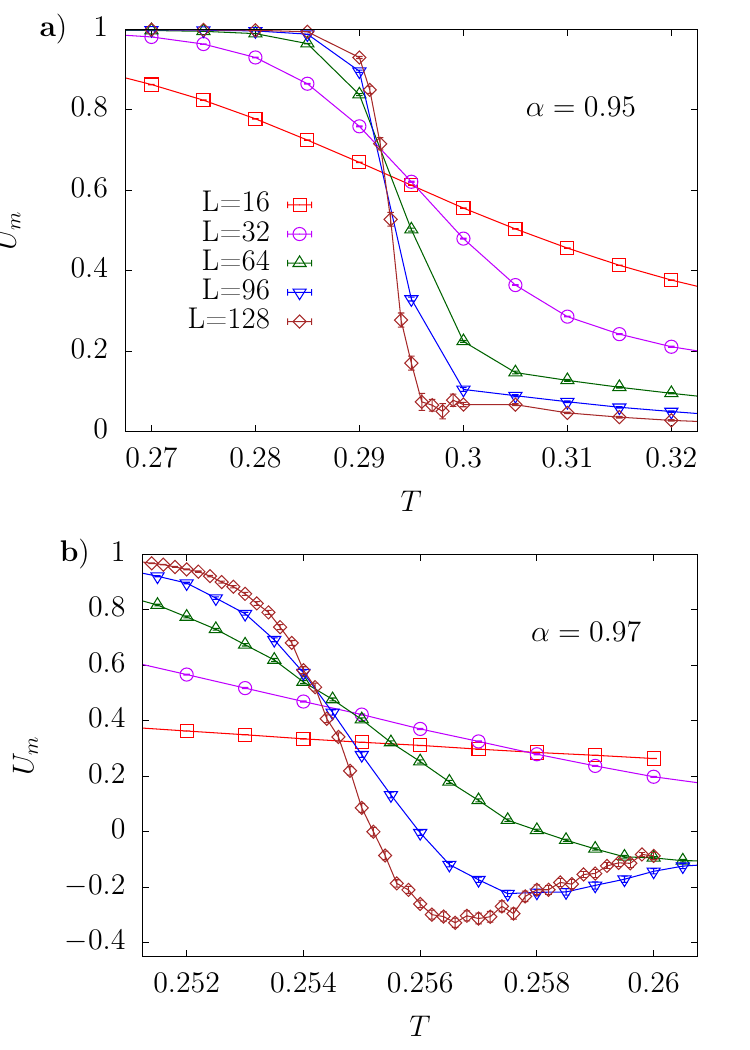}
\vskip 0.4cm
\caption{ Indication of first order transition at $\alpha=0.97$ from Binder cumulants $U_m$. 
	a) and b) show the Binder cumulants $U_m$ as a function of temperature $T$ for various system sizes $L$ at $\alpha=0.95,0.97$, respectively. 
	$U_m$ show negative peaks approaching larger negative values as the lattice size increases near the phase transition point 
	for $\alpha=0.97$, indicating a first-order phase transition, while staying positive in the whole temperature region for $\alpha=0.95$, typical for 
	a continuous transition.}
\label{um-0.95_0.97}
\end{figure}

Figure \ref{um-0.95_0.97} shows the Binder cumulants $U_m$ as a function of temperature $T$ for various 
system sizes $L$ at $\alpha=0.95$ and $0.97$, respectively. We can see that the Binder cumulants near the 
phase transition point are negative and tend to approach larger negative values as the lattice size increases for $\alpha=0.97$, indicating a first-order phase transition\cite{Vollmayr}.

\begin{figure}[t]
\centering
\setlength{\abovecaptionskip}{20pt}
\includegraphics[angle=0,width=0.5\textwidth]{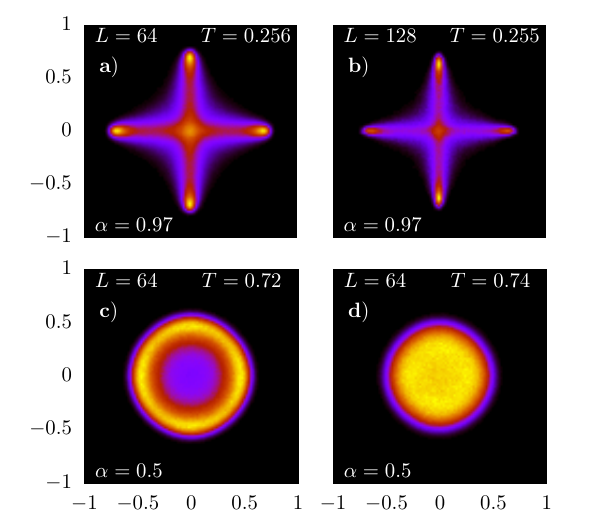}
\vskip 0.4cm
\caption{The 2D histogram of $\vec m$ at $\alpha=0.97$ and $\alpha=0.5$. 
	The $x$-axis represents $m_x$, while the $y$-axis represents $m_y$. The color in the figures represents the probability density, with brighter colors indicating higher probabilities. 
	a) and b) display the histograms at $\alpha=0.97$ for $L=64$ at $T=0.256$ and 128 at $T=0.255$, respectively. 
	The histogram reveals a coexistence of ordered and disordered phases, suggesting the possibility of a first-order phase transition. 
	c) and d) depict the histograms at $\alpha=0.5$ for $L=64$ at $T=0.72$ and $T=0.74$, respectively. There is no phase coexistence found, indicating a continuous phase transition, in contrast to the 
	histograms shown in a) and b).} 
\label{his2d}
\end{figure}

To further determine the property of the transition at $\alpha=0.97$, we calculated and compared the histograms of $\vec m$ for $\alpha=0.97$ and $\alpha=0.5$,
as shown in Fig.\ref{his2d}.
At $\alpha=0.97$, the histogram reveals a coexistence of ordered and disordered phases, suggesting the possibility of a first-order phase transition when $\alpha$ is close to $1$.

\begin{figure}[t]
\centering
\setlength{\abovecaptionskip}{20pt}
\includegraphics[angle=0,width=0.45\textwidth]{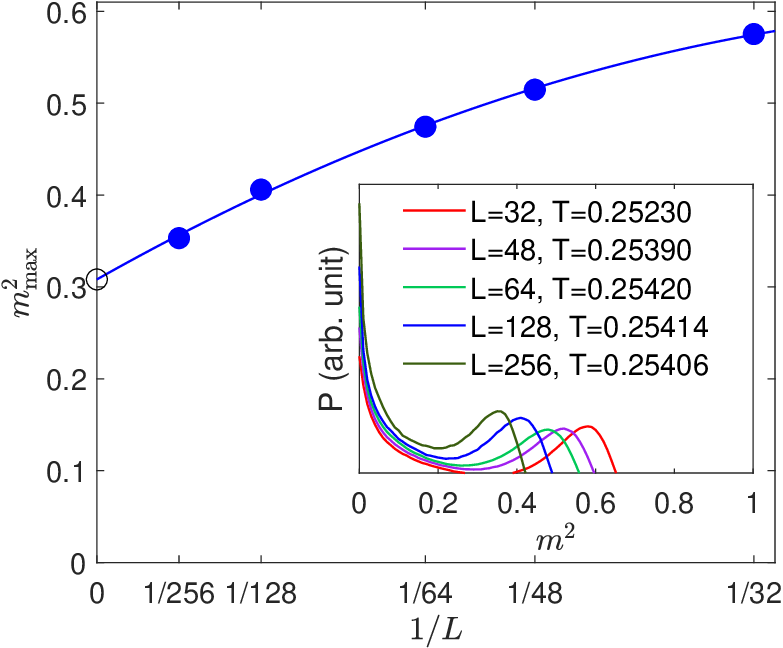}
\vskip 0.4cm
\caption{Finite-size analysis of the 1D histogram at $\alpha=0.97$. The inset shows the 1D $m^2$ histogram at the temperature where the probabilities of the ordered and disordered 
	phases are equal for various system sizes. In the main panel, the blue solid dots show $m^2_\mathrm{max}(L)$ having the maximum probability in the ordered phase 
	in the $m^2$ histogram. The blue line is a polynomial fit to the data, giving a value $m^2_\mathrm{max}(L=\infty)=0.308$. While we are not concerned the precise value of $m^2_\mathrm{max}(L=\infty)$, our analysis unambiguously indicates it is non-zero, allowing us to infer a first-order transition for this $\alpha$.   } 
\label{his1d}
\end{figure}

The 1D version of the $m^2$ histogram $P(m^2)$ at $\alpha=0.97$ is presented in Fig.\ref{his1d}.
Specifically, Fig.\ref{his1d}a) displays the $m^2$ histogram at the temperature where the probability of the ordered and disordered phases is equal for various system sizes. 
We can see that the probability density of the disordered phase tends to be infinity with the 
increase of $L$. 
Figure \ref{his1d}b) displays $m^2_\mathrm{max}(L)$, which 
has the maximum probability of the ordered phase in the $m^2$ histogram for system size $L$. Polynomial fit to the data up to the second order gives a finite value $m^2_\mathrm{max}(L=\infty)=0.308$.

\section{\label{sec:sum}Summary}

In summary, we have introduced a simple model $H_{\rm gCM}$ for a square lattice compass model with a tuning parameter $\alpha$, which has the generic feature of spin-lattice rotational symmetry but does not have any accidental subsystem symmetries. For $\alpha=0$ $H_{\rm gCM}$ becomes the XY model, and for $\alpha=1$, it is the compass model. Generically away from these limiting values, our model has an ordered phase at low temperatures, which breaks the spin-lattice rotational symmetry. The thermal phase transition between the ordered phase and the high-temperature disordered phase is found to be both continuous and first-order in different regions of the $\alpha$ parameter.  The region that is continuous displays Ashkin-Teller like criticality with continuously varying exponents as expected theoretically this line terminates at the four-state Potts point and then for larger $\alpha$ the transition turns first order. From our numerical studies, we find that the first-order transition line connects the 4-state Potts point to the well-known Ising criticality of the compass model at $\alpha=1$. This part of the phase diagram is intriguing and merits further investigation. 

Our gCM model joins a list of diverse statistical mechanics models which show this kind of novel AT critical behavior, including (but not limited to) the eight-vertex model~\cite{Baxter}, the XY model with four fold anisotropy~\cite{Jose-clock}, the J1-J2 Ising model~\cite{Jin2013},  models of mixtures of dimers with momomers and squares~\cite{ramolaAT, Morita_2023} and thermal transition of VBS order~\cite{vbsthermal}. 

In future work, it would be of interest to study the phase diagrams of three-dimensional versions of our model, as well as the phase diagram of the quantum version of $H_{\rm gCM}$.

\begin{acknowledgments}
We thank G. Murthy and A. Sandvik for their valuable discussions. This work was supported by the National Natural Science Foundation of China under Grant No. 12175015 and by NSF DMR-2026947.

\end{acknowledgments}

\clearpage

 \bibliography{compass.bib}

\end{document}